\documentclass[5p,twocolumn]{elsarticle}

\usepackage{amsmath}
\usepackage{amssymb}
\usepackage{mathrsfs} 
\usepackage{bm}

\usepackage{algorithm}
\usepackage[noend]{algpseudocode}

\usepackage{ulem}

\usepackage{phys}
\usepackage{mystyle}

\journal{Physica B}

\begin{document}

\begin{frontmatter}



\title{A regressor-based hysteresis formulation for the magnetic characterisation of low carbon steels}

\author[CEA]{A. Skarlatos}
\author[CEIT,UNav]{A. Mart\'inez-de-Guerenu}
\author[CEA]{R. Miorelli}
\author[CEIT,UNav]{A. Lasaosa}
\author[CEA]{C. Reboud}
%
\address[CEA]{CEA, LIST, CEA Saclay, Gif-sur-Yvette F-91191, France}
\address[CEIT]{CEIT, Manuel Lardizabal 15, 20018 Donostia / San Sebastián, Spain}
\address[UNav]{Universidad de Navarra, Tecnun, Manuel Lardizabal 13, 20018 Donostia / San Sebastián, Spain}


\begin{abstract}
In this work, two different parametric hysteresis models, the Jiles-Atherton model and the 
Mel'gui relation, have been combined to form a more general hysteresis operator, suitable for 
the description of families of experimental B(H) curves obtained for low carbon (LC) steel 
specimens after isothermal annealing at different temperatures and times. As it has been 
demonstrated in a number of previous studies, characteristic values of steel hysteresis 
curves can be used as very efficient identifiers for the monitoring of the different 
metallurgical transformations that take place during the annealing, such as recovery and 
recrystallisation processes. It is thus important from a practical point to be able to 
reproduce the experimental curves obtained under different conditions, as precisely as 
possible, in order to proceed to the samples characterisation. Hybridisation of the two 
aforementioned models demonstrated satisfactory results for the reproduction of all considered 
curves obtained under the different considered annealing conditions.
\end{abstract}


\end{frontmatter}


\section{Introduction}

Different parametric models have been proposed in the literature for the modelling of magnetic
hysteresis, yet none of the existing models proves to be, to the authors knowledge, universal. 
In fact, a specific model may be well adapted for the description of a material family but it 
can perform poorly for others. This drawback may be acceptable when one is interested in a 
given material. They exist, however, situations where a continuous variation of the material 
properties may produce hysteresis loops with broad range of features that cannot be captured 
by a single model.

This situation seems to be the case when we examine the magnetic hysteresis loops of cold 
rolled extra low carbon steels which are subjected to thermal annealing treatments. It turns 
out that the hysteresis loops of samples subjected to isothermal annealing at different 
temperatures and for different holding times, span a range of loop shapes, which is 
consequence of the different microstructural transformations that take place during the 
annealing processes.

Previous works have revealed that certain characteristics of the hysteresis loops, like the
coercive field, the remanent magnetisation or the hysteresis losses correlate well with 
microstructural parameters like the grain size and the dislocation density and they can thus
be used as identifiers for the evolution monitoring of transformation procedures such as the
recovery and the recrystalisation \cite{martinezdeguerenu_actamater04a, 
martinezdeguerenu_jmm07}. The thereupon presented results justify the need for numerical 
hysteresis models able of predicting the basic features of the hysteresis curves for the
whole range of the family. 

Numerical experimentation with existing parametric hysteresis models like the Jiles-Atherton 
model \citep{jiles_jmmm86} or the Mel'gui model \citep{melgui_defekt87} revealed that the
identification of the two models using a standard iterative optimisation procedure 
yields different results for the ensemble of the experimental curves obtained via the 
procedure described in  \cite{martinezdeguerenu_actamater04a, martinezdeguerenu_jmm07}.
The reasons for these discrepancies may be attributed to the limitations of the models or even 
to the optimisation procedure itself, in the sense that regions with non-physical output for 
the model may be visited during the exploration of the input space.

To improve the precision of the representation, a mixing of the two models is proposed in this
work. The main idea consists in sampling a common input space, consisting of characteristic
hysteresis features like the ones mentioned above, and apply the best approach for each of the
points of the input space based on a series of criteria concerning the form of the output 
curves. The sampling points are then used as training set for a  Gaussian process regressor, 
which replaces the physical model thus becoming a generic hysteresis operator. In this way one 
can assure smooth variations across the input spaces and avoid "holes" with non-physical 
outputs. Although the physical parametric models considered here are the Jiles-Atherton and 
the Mel'gui model, the proposed approach is model independent making other combinations 
possible.

The paper is organised as follows. In the first section the Jiles-Atherton and the Mel'gui 
model and some representative results of their identification for two different hysteresis 
loops are briefly presented. In the second section the hysteresis operator is 
reformulated in a more formal way, and the algorithm for the construction of the regressor 
model is presented. A brief presentation of the mathematical model of the Gaussian process 
regressor is given in the next paragraph. Details on the experimental procedure for the 
production of the material curves are provided in a separate section. The results of the 
approach for the reproduction of some representative experimental curves are presented at the 
last section.

\section{Identification of the Jiles-Atherton and the Mel'gui model}

In the Jiles-Atherton model \citep{jiles_jmmm86} the total magnetisation of the material $M$ 
is understood as the combined effect of two contributions: a reversible $M_{rev}$, related to 
the domain walls motion and an irreversible $M_{irr}$ one, which is basically the result of 
the domain bulging. The two components are given by the equations
%
\begin{equation}
M_{rev} = c\left(M_{an} - M_{irr}\right)
\end{equation}
and
%
\begin{equation}
\frac{M_{irr}}{dH} = 
\frac{M_{an} - M_{irr}}{k\delta/\permb_0 + \alpha\left(M_{an} - M_{irr}\right)}
\end{equation}
where $c$ is a proportionality coefficient with $c\in[0, 1]$, $\permb_0$ stands for the 
magnetic permeability of the free space, and $k$ and $\alpha$ are material constants related 
to the domain wall pinning and the interdomain coupling, respectively. $\delta=\pm 1$ is 
essentially a numerical flag, which distinguishes between the descending and ascending 
branches. The $M_{an}$ term stands for the anhysteretic material curve, calculated via the 
implicit relation
%
\begin{equation}
M_{an} = M_s L\fun{\frac{H + a M_{an}}{a}}
\end{equation}
with $L\fun{x} = \coth\fun{x} - 1/x$ being the Langevin function. $M_s$ gives the 
magnetisation at saturation, and $a$ is another material parameter related to the domains 
density.

In the Mel'gui model, the magnetisation is approximated by the following closed-form relation
%
\begin{align}
M &= 
\chi_{in}\frac{H_c^2 H}{H^2+H_c^2}
+\delta
\frac{M_s}{\pi}
\frac{H_m^2}{H_m^2+b H_c^2}
\nonumber\\
&
\times\left\{
2\arctan\fun{\frac{H_c + \delta H}{H_0}}
\right.
\nonumber\\
&-
\left.
\left[
\arctan\fun{\frac{H_c + H_m}{H_0}} +
\arctan\fun{\frac{H_c - H_m}{H_0}}
\right]
\right\}
\label{eq:Melgui}
\end{align}
with the coefficients $H_0$ and $b$ being defined as
\begin{equation}
H_0 = \frac{H_c}{\tan\fun{\pi M_r/2M_s}}
\end{equation}
and 
\begin{equation}
b = \frac{M_s}{\pi} \frac{\arctan\fun{2H_c/H_0}}{M_c - \chi_{in}H_c/2}.
\end{equation}
In the above equations, $H_c$, $M_s$, $M_r$ stand for the coercive field, the magnetisation at
the saturation and the remanent magnetisation respectively, $\chi_{in}$ is the initial 
susceptibility (susceptibility of the demagnetised state) and $M_c$ the point of the first 
magnetisation curve at field intensity equal to the the coercive field.

The basic difference of the two models consists in the fact that whereas the magnetisation in
the Jiles-Atherton model is obtained via the solution of an ordinary differential equation, 
in the Mel'gui model it is evaluated directly via a closed form relation, which makes the
latter much faster. The domain of application is not the same either. Thus, the Jiles-Atherton
model can theoretically be applied for an arbitrary excitation cycle (although the accuracy
of the model for the description of minor loops is questionable), whereas the Mel'gui relation
is only applicable to symmetrical periodic excitations. The input domain of the two models is
also very different with four of the five parameters of the Jiles-Atherton model being 
internal tuning parameters that need to be determined via the so-called model identification
procedure. The Mel'gui model, on the contrary, is parametrised via characteristic values of 
the major loop and the initial magnetisation curve, which makes the model directly applicable.
Nevertheless, numerical experimentation reveals that the parameter estimation via 
identification (that is via the solution of an optimisation problem) yields better results for
this model as well.

\begin{figure}[h]
\centering
\includegraphics[width=9.0cm]{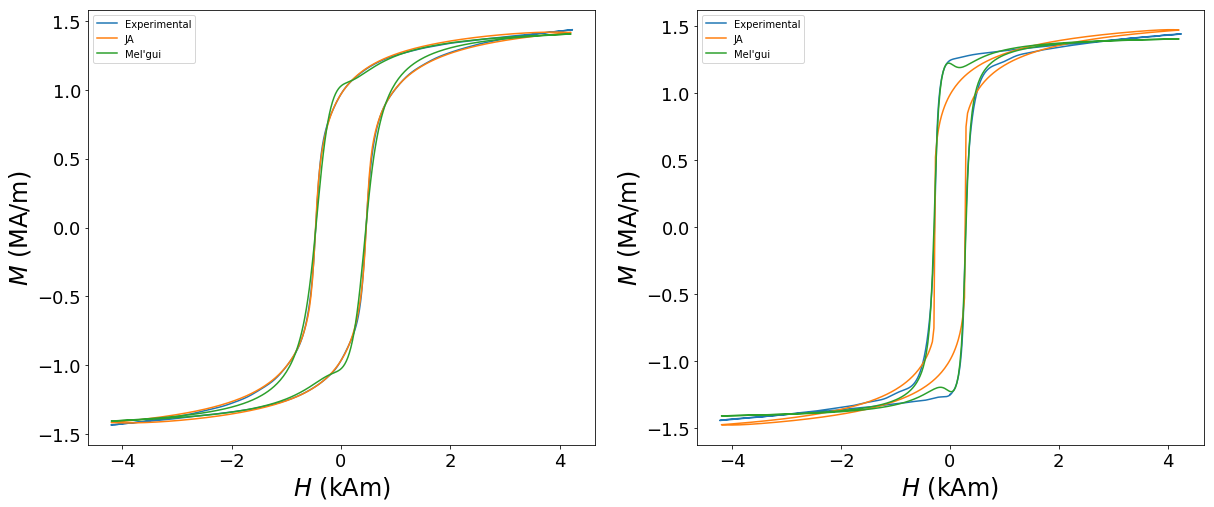} \\
\caption{Hysteresis curves calculated using the Jiles-Atherton and Mel'gui model after 
identification with different experimental data. The experimental curves used for the
identification are given for comparison.}
\label{fig:ident_example}
\end{figure}

An example of the two models identification using two different experimental curves with 
different coercivity is shown in \figref{fig:ident_example}. The models have been identified
using a standard optimisation approach, where the $L_2$ norm of the residual between the 
theoretical and the experimental curve is minimized. It turns out that for the given 
minimisation procedure the Jiles-Atherton model performs slightly better than the Mel'gui
model for the first curve (corresponding to a harder material) whereas the tendency is 
inverted for the second curve. As far as the second result is concerned, a possible 
explanation for the poor performance of the Jiles-Atherton model could be that the utilised 
optimisation algorithm by trying to adapt the curve to the very steep slope around the 
coercive field does not succeed to fulfil all the rest of the form constrains at the same 
time. One must also not exclude the fact that regions with non-physical model solutions may 
be visited during the exploration of the input space, which destabilises the procedure and 
may lead to sub-optimal results.

\section{Construction of the regressor model}

\subsection{Hysteresis description in different input spaces}

The magnetisation for a scalar hysteretic magnetic material admits the general expression
%
\begin{equation}
M = M\fun{H;H_{0},H_{-1},\ldots}
\label{eq:hystgen}
\end{equation} 
where $H$ is the applied magnetic field at time $t$, and $H_{-r},\; r=0,1,\ldots$ stand for 
the field reversal points (i.e. the points where the time derivative of $H$ changes sign) at 
previous instances $t_{-r} < t$. The value of the magnetic field $H$ together with the 
ensemble of reversal points $H_{-r}$ fully determine the state of the material taking full
account of the excitation history.

Let us consider the hysteresis relation \eqref{eq:hystgen} along an observation window
$t\in[0, T]$, which is sampled using a homogeneous temporal grid $t_i = i T/N$, with 
$i=0,...,N$. Assuming that the magnetic field values at the sample points of the observation 
window are known, the magnetic field vector $H_i = H\fun{t_i}$ fully determines the 
magnetisation by application of \eqref{eq:hystgen} since $H_i$ contains all the information
of the field history (i.e. the reversal points $H_{-r}$).

We seek to establish the following mapping $\Hfield \rightarrow \Magn$ with
%
\begin{equation}
\vect{M} = M\fun{\vect{H}}.
\label{eq:MHmapping}
\end{equation}
Notice that $\vect{H} = [H_0,\ldots, H_N]^T$ and $\vect{M} = [M_0,\ldots, M_N]^T$ stand here 
for column vectors comprising the values of the discretised magnetic field and magnetisation 
samples inside the observation window, which should not be confused with the vector 
counterparts of the two field variables.

We assume that a parametric model is used for the numerical evaluation of 
\eqref{eq:MHmapping}, which means that each pair $(\Hfield, \Magn)$  is associated to a set
of parameters used for the tuning of the model of choice. Let $\vect{p}$ be the vector 
containing the values of the model parameters. One can formally write
%
\begin{equation}
\Magn = M_{\Hfield}\fun{\vect{p}},
\end{equation}
where the index $\vect{H}$ is used to denote that the above mapping is valid for a given 
magnetic field discretisation $\Hfield$. For the sake of notational simplicity, the model 
dependence on the $\Hfield$ vector will be implied in the rest of the paper. 

Our objective is to combine two (or more) different parametric models $M_a$ and $M_b$ into a 
common hysteresis operator, by either switching between the two models, interpolate between 
points calculated with the two models or by performing a weighted sum of their outputs. The 
three operations can be expressed in terms of a weighted sum 
\begin{equation}
\Magn = w_a\fun{\vect{p}_a} M_a\fun{\vect{p}_a} + w_b\fun{\vect{p}_b} M_b\fun{\vect{p}_b},
\label{eq:modelcomb}
\end{equation}
where $w_a\fun{\vect{p}_a}$ and $w_b\fun{\vect{p}_b}$ stand for the corresponding weighting 
coefficients, whose value vary between 0 and 1 depending on the given parameter combination.
In order to build that operator, the input vectors of the two consisting models $\vect{p}_a$ 
and $\vect{p}_b$ must be expressed in a new parametric space, meaningful for both models.

The most straight-forward choice for this new space is to pick a number of hysteresis 
characteristic points or slopes, which are common features of all hysteresis curves and 
independent of the model details. A possible input set (among others) is the 
$\vect{p}_c = (M_s, H_c, M_r, W_h, \chi_r)$, where $M_s$ is the magnetisation at saturation, 
$H_c$ the coercive field, $M_r$ the remanent magnetisation, $W_h$ the hysteresis losses and 
$\chi_r$ the susceptibility at the remanence. We can rewrite then \eqref{eq:modelcomb} 
formally in the following way
\begin{equation}
\Magn = 
w_a\bfun{\vect{p}_a\fun{\vect{p}_c}} M_a\bfun{\vect{p}_a\fun{\vect{p}_c}} + 
w_b\bfun{\vect{p}_b\fun{\vect{p}_c}} M_b\bfun{\vect{p}_b\fun{\vect{p}_c}}.
\end{equation}

In order to proceed to the numerical evaluation of the scheme, one has to calculate the 
weighting coefficients $w_a\fun{\vect{p}_c}$ and $w_b\fun{\vect{p}_c}$ as functions of the 
new coordinates and to carry out the coordinate transformations $\vect{p}_a\fun{\vect{p}_c}$ 
and $\vect{p}_b\fun{\vect{p}_c}$. The algorithm of the regressor model is summarized in form 
of pseudo-code in the following table. 
\begin{algorithm}
\mbox{Off-line phase:}
\begin{algorithmic}[1]
\State Define the domains $\vect{P}_a$, $\vect{P}_b$, with $\vect{p}_a \in \vect{P}_a$, 
$\vect{p}_b \in \vect{P}_b$ 
\State Random sampling of $\vect{p}_a$ and $\vect{p}_b$: 
get $\vect{p}_{a,i}$ and $\vect{p}_{b,i}$
\For{$i=1\ldots N_s$}
\State Evaluate $M_a\fun{\vect{p}_{a,i}}$, $M_b\fun{\vect{p}_{b,i}}$
\If{$M_a\fun{\vect{p}_{a,i}}$, $M_b\fun{\vect{p}_{b,i}}$ non-physical}
	\State Discard $i$
\Else
	\State Evaluate the new coefficients 
	$\vect{p}_{c,i} = \vect{p}_{c}\fun{\vect{p}_{a_i}, \vect{p}_{b_i}}$
\EndIf
\EndFor
\State Train the regressor $\mathscr{M}\fun{\vect{p}_{c,i}} = 
w_a\fun{\vect{p}_{a,i}} M_a\fun{\vect{p}_{a,i}} + w_b\fun{\vect{p}_{b,i}} M_b\fun{\vect{p}_{b,i}} = \Magn_i$, 
$\forall i$
\end{algorithmic}
\mbox{On-line phase:}
\begin{algorithmic}[1]
\State Evaluate $\mathscr{M}\fun{\vect{p}_c}$, with $\vect{p}_c \in \vect{P}_c$
\end{algorithmic}
\caption{Regressor model}
\end{algorithm}

\begin{figure}[h]
\centering
\includegraphics[width=9.0cm]{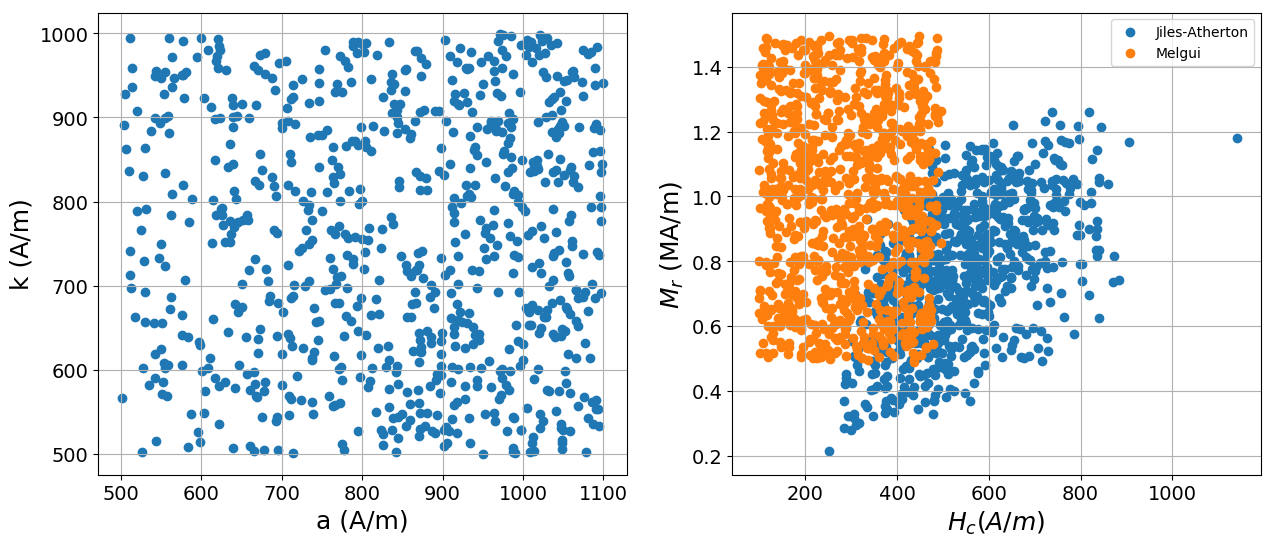} \\
\begin{tabular}{cc}
\small{(a)} \hspace{3.0cm} & \small{(b)}
\end{tabular}
\caption{Coordinate transformation. (a) Random sampling of the $(a, k)$ subspace of the 
Jiles-Atherton native parameter space. (b) Sampling of the characteristics subspace 
$H_c - M_r$. With blue dots are marked the sampling points of the Jiles-Atherton model,
given on the left figure, as they are transformed to the $H_c - M_r$ subspace. The orange 
points correspond to the sampling points of the Mel'gui model. Notice that the given subspace
is a native subspace of the Mel'gui model (both $H_c$ and $M_r$ belong to its inputs).}
\label{fig:subspace}
\end{figure}

For the case of the Jiles-Atherton and the Mel'gui model, which are the models of the choice 
in this work the two input vectors are $\vect{p}_{a} \equiv \vect{p}_{JA} = (M_s, k, a, \alpha, c)$
and $\vect{p}_{b} \equiv \vect{p}_{Mel} = (M_s, H_c, M_r, M_c, \chi_{in})$. A representative 
sampling of a $\vect{p}_c$ subspace is shown in \figref{fig:subspace}, where the corresponding 
mapping from the Jiles-Atherton native parameter space $\vect{p}_{JA} \mapsto \vect{p}_c$ is 
also illustrated .

\subsection{Hysteresis model based on Gaussian process regression}

In the previous paragraph, a strategy of replacing the physical hysteresis model by a 
metamodel (regressor in this case) built upon a number of pre-calculated data sets was 
presented. Very recently, different regression approaches based on metamodels have been 
studied in the literature of non destructive testing showing good accuracy in approximating 
unseen experimental data \citep{salucci_transgeoscirs16, ahmed_ndte19}. Here we shall examine 
the case of the Gaussian process regression as it applies for the representation of hysteresis 
data.

Let us consider the hysteresis model sampled over a set of 
$\mathbf{P}=\left[\mathbf{p}_{1},...,\mathbf{p}_{N}\right]^{\mathrm{T}}$
parameters along with the respective \textendash for sake of simplicity\textendash{}
scalar outputs $\Magn\left(\mathbf{p}_{1}\right),...,\Magn\left(\mathbf{p}_{N}\right)$.
Where the $i$-th entry $\mathbf{p}_{i}\in\mathbb{R}^{1\times P}$.
Let us now consider the aforementioned outputs as a stochastic process
where each entry is given by a random vector defined as 
$\mathscr{M}=\left(\mathscr{M}\left(\mathbf{p}_{1}\right),...,
\mathscr{M}\left(\mathbf{p}_{N}\right)\right)$,
where $\mathscr{M}$ stands for a random vector realisation. One can express the correlation
between random variables as an exponential function with power equal to $2$ which corresponds 
to the so-called Gaussian kernel function. Therefore, the correlation function calculated on 
the hysteresis model turns into 
\begin{equation}
\phi\left({p}_{i},{p}_{j}\right)=
\exp\left(-\sum_{p=1}^{P}\theta_{j}\left|\mathbf{p}_{i}-\mathbf{p}_{j}\right|^{2}\right),
\label{eq:correlation_func}
\end{equation}
where $\theta_{j}$ is a hyper-parameter to be estimated via maximum
likelihood estimation or cross validation methods. Starting from the
definition of the correlation function between two random variables,
one can show that the covariance matrix between these variables is
given as $\mathrm{Cov}\left(\mathscr{M},\mathscr{M}\right)=
\sigma_{\mathscr{M}}^{2}\mathbf{\Phi}\left(\mathscr{M}\right)$
\citep{forrestereng} where $\mathbf{\Phi}$ being the $N\times N$
correlation matrix and $\sigma_{\mathscr{M}}$ is the
standard deviation of $\mathscr{M}$. That is, the stochastic
model considers the correlation between sampled data that can be accounted
via a specific correlation model described by $\mathbf{\Phi}$. Therefore, the model depends on 
the distances between the considered sampled points. The Ordinary Kriging (OK) prediction 
($\widehat{\mathscr{M}}\left(\cdot\right)$) on a new entry $\mathbf{p}_{N+1}$ is obtained by 
\[
\widehat{\mathscr{M}}\left(\mathbf{p}_{N+1}\right)=
\sum_{p=1}^{P}\lambda_{i}\mathscr{M}\left(\mathbf{p}_{i}\right),
\]
where $\lambda_{i}$ are the kriging coefficients to be estimated 
\citep{villemonteix_jglobopt08}. Thus, accounting the unbiasedness of the predictions (i.e., 
$\widehat{\mathscr{M}}\left(\mathbf{p}_{i}\right)=\mathscr{M}\left(\mathbf{p}_{i}\right)$)
leads to the following condition 
\[
\sum_{p=1}^{P}\lambda_{i}=1,
\]
which translates into the best linear unbiased prediction obtained by the minimisation of 
the mean square prediction error (i.e., the squared expectation) as 
\citep{villemonteix_jglobopt08}
\begin{equation}
\underset{\mathrm{\lambda}_{i}}{\mathrm{min}}\,\,\mathbb{E}
\left(\left[\widehat{\mathscr{M}}\left(\mathbf{p}_{N+1}\right)-
\sum_{p=1}^{P}\lambda_{i}\mathscr{M}\left(\mathbf{p}_{i}\right)\right]\right)^{2}
\,\,\,\,\,s.t.\,\,\sum_{i=1}^{P}\lambda_{i}=1.
\label{eq:OK_minimisation}
\end{equation}
The kriging coefficients in (\ref{eq:OK_minimisation}) can be obtained
by applying the method of Lagrange multipliers. The optimisation boils
down to the following matrix system of equations
\begin{equation}
\left[\begin{array}{cc}
\mathbf{\Phi} & \mathbf{U}\\
\mathbf{U}^{\mathrm{T}} & \mathbf{0}
\end{array}\right]\left[\begin{array}{c}
\mathbf{\lambda}\\
\mu
\end{array}\right]=\left[\begin{array}{c}
\bm{\phi}\\
\mathbf{u}
\end{array}\right],\label{eq:OK_system}
\end{equation}
where in (\ref{eq:OK_system}),
$\mathbf{\Phi}=\left[\phi\left(\mathbf{x}_{i},\mathbf{x}_{j}\right)\right]\, \forall i,j\in1,...,N$
and $\bm{\phi}=\left[\phi\left(\mathbf{x}_{1},\mathbf{x}_{N+1}\right),\phi\left(\mathbf{x}_{2},
\mathbf{x}_{N+1}\right),...,\phi\left(\mathbf{x}_{N},\mathbf{x}_{N+1}\right)\right]$ is defined 
as provided in (\ref{eq:correlation_func}).
Moreover,
$\mathbf{U}=\left[\mathbf{u}\left(\mathbf{p}_{1}\right)^{\mathrm{T}},...,
\mathbf{u}\left(\mathbf{p}_{N}\right)^{\mathrm{T}}\right]^{\mathrm{T}}$
with $\mathbf{u}\left(\mathbf{p}_{i}\right)$ being a lower-order monomials
(typically it does not exceed the degree of two), $\mathbf{0}$ is a zeros matrix and $\mu$ 
represents the vector of $l<N$ the Lagrange multipliers.

\section{Application to microstructure monitoring during annealing treatment}
\label{}

\subsection{Sample preparation and hysteresis measurements}

\begin{figure}[h]
\centering
\includegraphics[width=8.0cm]{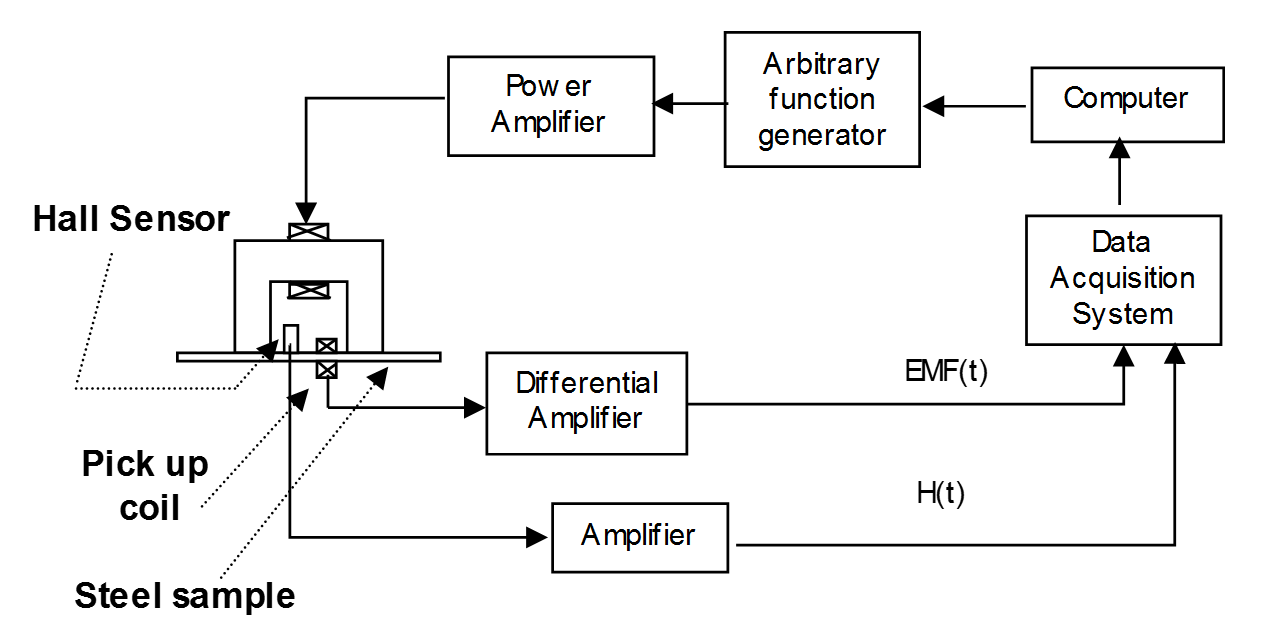} \\
\caption{Experimental setup for the acquisition of the B-H curves.}
\label{fig:expsetup}
\end{figure}

Experimental magnetic measurements on cold rolled samples that had been annealed at low 
temperatures (300-500~$^\circ$C) in order to promote recovery without interaction with 
recrystallisation and at 600~$^\circ$C to induce recrystallisation were used in this study 
\cite{martinezdeguerenu_actamater04a, martinezdeguerenu_jmm07}. The original samples were from 
extra low carbon steel, with composition 0.03\%C-0.19\%Mn-0.13\%Al-0.0035\%N-0.012\%P-0.01\%Si, 
that had been industrially produced and cold rolled to a final thickness of 0.3 mm through a 
reduction of 84\% \cite{martinezdeguerenu_actamater04a}. Near saturation major magnetic B-H 
hysteresis loop determination was made using a single sheet tester system available at CEIT 
\citep{soto_transim09} at 1 Hz, with maximum magnetic field strengths applied of about 
4100 A/m. The schematic diagram of the B-H measurement system is shown in 
\figref{fig:expsetup}.
%

The external magnetic field was produced by a magnetic yoke composed of a 200-turn coil wound 
around a U-shaped magnetic laminated core. The excitation signal was generated by a sinusoidal 
magnetizing current produced by a programmable function generator connected to a power 
amplifier. The induced electromotive force in an encircling coil wound around the samples and 
the tangential magnetic field strength measured using a Hall probe placed at the surface of 
the samples, were acquired using a NI data acquisition system. 

Four major hysteresis loops were recorded for each measurement applying a sinusoidal magnetic 
field strength of about 4.1~kA/m at 1 Hz, which was sufficiently high to reach near saturation 
state of the measured samples. These were demagnetized prior to each test. The sampling 
frequency used was 5 kHz.

Recovery involves both the annihilation of dislocations and their rearrangement into low 
energy configurations. Recrystallisation leads to the suppression of dislocations by the 
nucleation of defect free volumes and by the migration through the material of the 
recrystallisation front, resulting in a new grain structure with a low dislocation density. 
Previous studies [1] showed that coercive field measurements can be satisfactorily employed to 
monitor recovery during low temperature annealing, during which the grain structure remains 
constant and microstructural changes only occur in the cold rolling dislocation substructure 
inside the grains. During recrystallisation both the effect of the reduction of the 
dislocation density and the change in the grain size have to be taken into account.

\subsection{Identification of the regressor model using experimental data for different annealing conditions}

\begin{figure}[h]
\centering
\includegraphics[width=8.0cm]{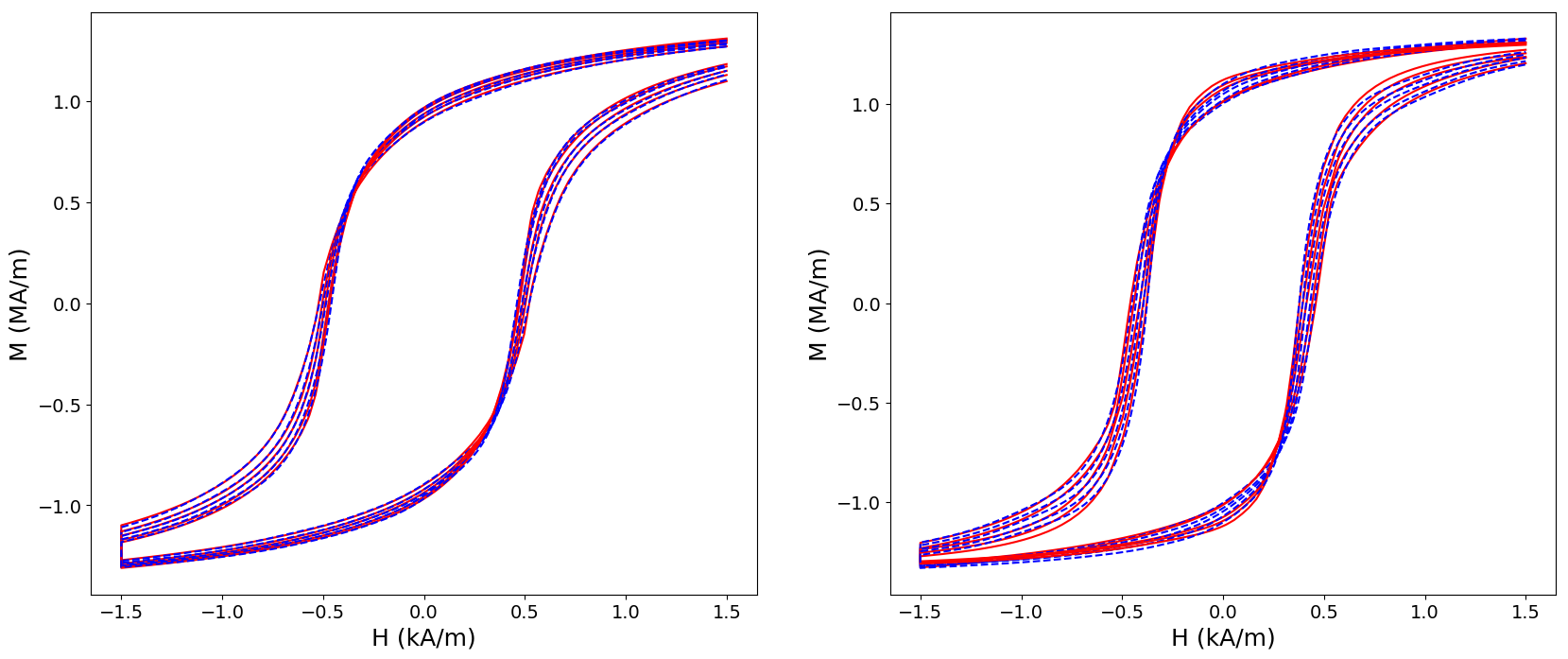} \\
\begin{tabular}{cc}
\small{(a)} \hspace{4.0cm} & \small{(b)}
\end{tabular}
\includegraphics[width=8.0cm]{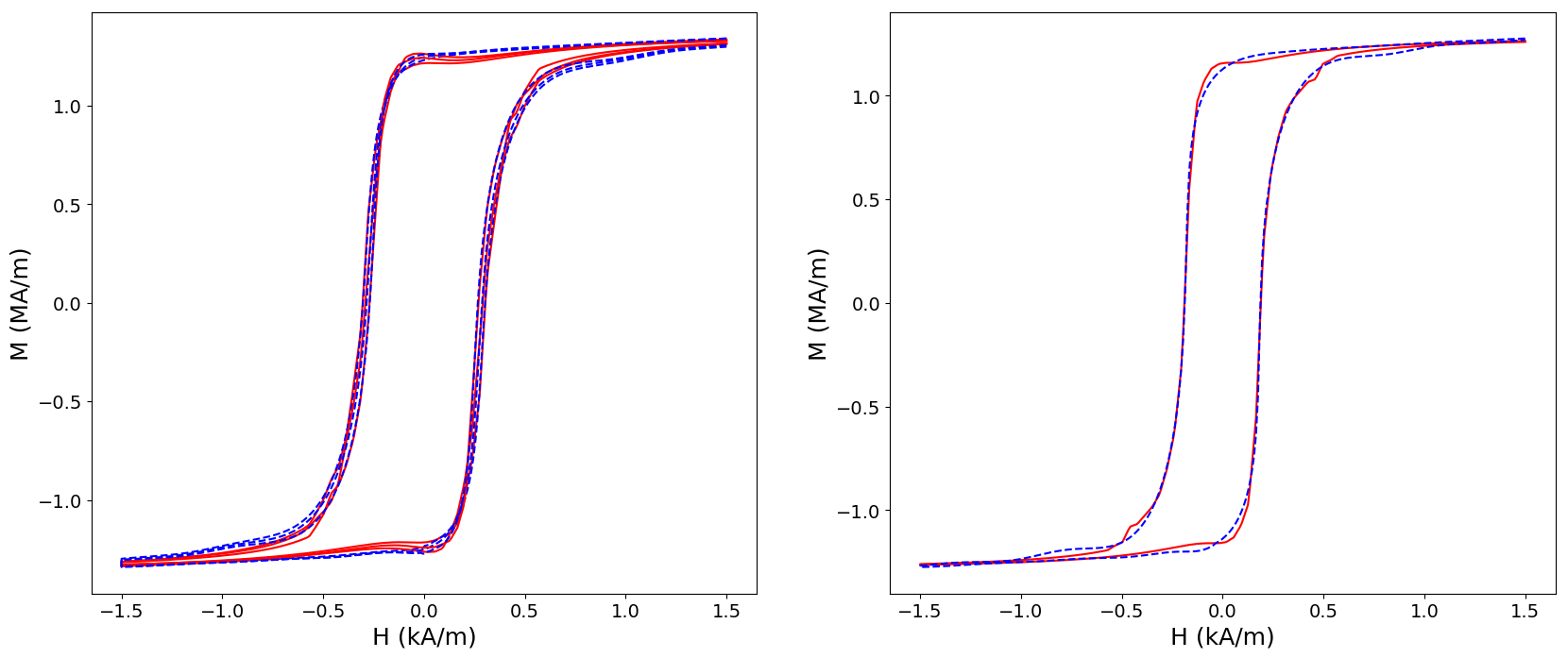} \\
\begin{tabular}{cc}
\small{(c)} \hspace{4.0cm} & \small{(d)}
\end{tabular}
\caption{Experimental vs. simulated curves for the annealing conditions of 
\tabref{tab:anneal_cond}. (a) 300~$^\circ$C, (b) 400~$^\circ$C, (c) 500~$^\circ$C and 
(d) 600~$^\circ$C. The experimental curves are drawn with dotted curves whereas the solid 
curves are the calculated hysteresis loops.}
\label{fig:res}
\end{figure}

The proposed approach has been applied for the reproduction of the experimental curves 
obtained from a cold-rolled (CR) low carbon (LC) steel sheets annealed at four different 
temperatures and for different holding times. Four temperatures are considered, namely 
300~$^\circ$C, 400~$^\circ$C, 500~$^\circ$C, and 600~$^\circ$C. The annealing times for the 
four temperatures are given in \tabref{tab:anneal_cond}.
\begin{table}
\begin{tabular}{l|l}
Temperature & Annealing time \\
\hline
300~$^\circ$C & 51~s, 4~min, 12~min, 36~min, 1.2~h \\
400~$^\circ$C & 11~s, 51~s, 4~min, 12~min, 36~min, 1.2~h \\
500~$^\circ$C & 51~s, 4~min, 12~min, 36~min \\
600~$^\circ$C & 51~s \\
\end{tabular}
\caption{Annealing conditions for the cold rolled steel samples.}
\label{tab:anneal_cond}
\end{table}
The predicted simulation curves are compared with the experimental ones for the four annealing 
temperatures in \figref{fig:res}. It should be noticed at this point that a classical 
iterative Jiles-Atherton model identifications works very well for the curves of the lowest 
two temperatures, but it does not succeed to reproduce correctly the steeper ones obtained at 
500~$^\circ$C and 600~$^\circ$C. This tendency is inverted in the case of the Mel'gui model, 
which performs better for the steeper curves obtained at higher temperatures. A typical set of 
hysteresis characteristic features obtained for a selected set of four curves, one for each 
annealing temperature is given in \tabref{tab:optvalues}.
%

The two datasets used for the construction of the regressor, i.e. the datasets for the 
Jiles-Atherton and the Mel'gui model, contained 1803 and 1886 curves respectively. Each
curve has been sampled using 600 points, which results in the storage of two databases
of approximately 20~MB each. The construction of the regressor, which consists the most
time consuming step, was carried out in 178~s. Once the regressor had been constructed,
the identification of the four curves demonstrated in \figref{fig:res} was carried out
in 3.18~s. The identification was based on five-parameters optimisation using 
the numpy python library implementation of the differential evolution algorithm. 
The average number of cost function evaluations needed to achieve the optimum was 2628, 
which results in an overhead per evaluation of circa 0.001~s. For the 
sake of comparison, it can be stressed out that the time demanded for a five-parameters 
optimisation using direct evaluation of the Jiles-Atherton model was of the 
order of 130~s. Both calculations were carried out in a standard DELL Precision T1700 
workstation with an Intel Xeon CPU E3-1241 v3 and 16 Gbytes RAM.
This drastic reduction of the computational time can be attributed to 
the very fast evaluation of each candidate curve via the regression approach (in fact
we interpolate in a set of pre-calculated curves), and constitutes one of the major
advantages of the proposed approach.

\begin{figure}[h]
\centering
\includegraphics[width=8.0cm]{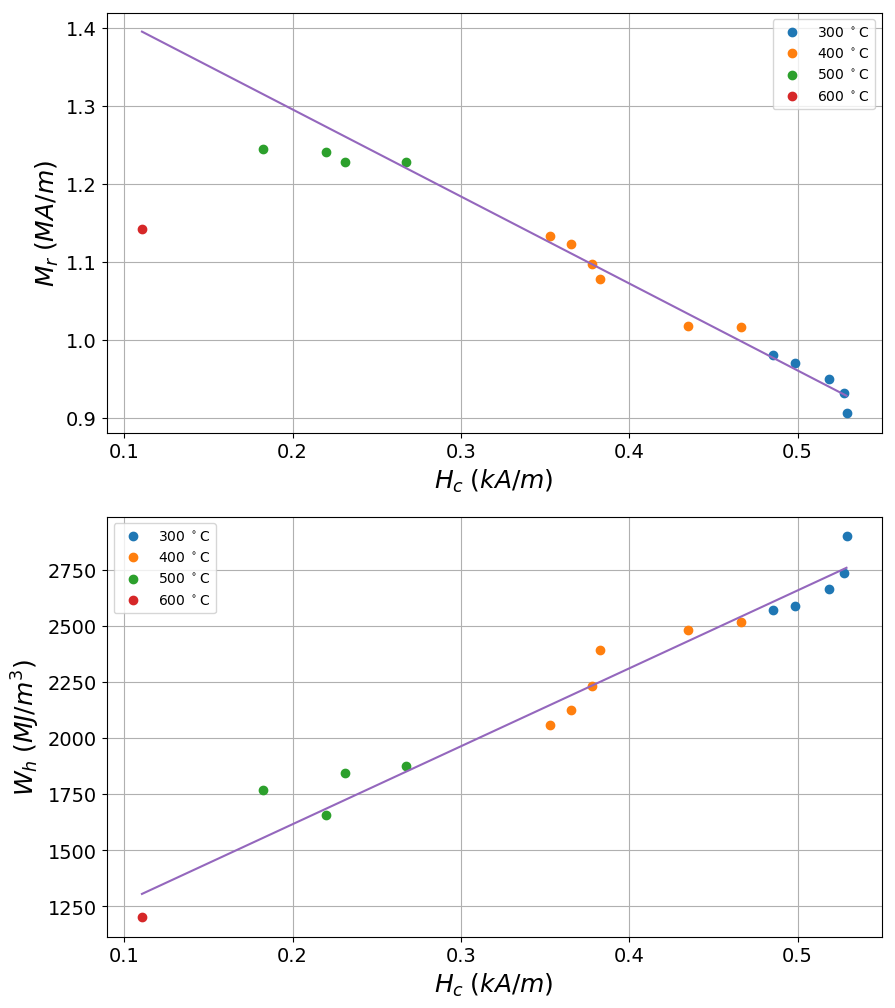}
\caption{Correlation of the remanent magnetisation and the hysteresis losses with the coercive
field.  a) $H_c$ vs. $M_r$ , (b) $H_c$ vs. $W_h$.}
\label{fig:correl}
\end{figure}

The $H_c$ vs. $M_r$ and $H_c$ vs. $W_h$ correlation plots deliver important information 
about the metalurgical transformation, which makes them a further test for the performance
of the regressor. The correlation plots for the calculated curves are shown in 
\figref{fig:correl}. It is important to notice that both pairs demonstrate a linear 
correlation for the first three temperatures, which is the trend observed using the 
experimental data \citep{martinezdeguerenu_actamater04a, martinezdeguerenu_jmm07}. A slight 
deviation from the strict linear law is observed at the  $H_c$--$W_h$ plot for the last 
three points (longer annealing) of the third temperature. The point corresponding to the 
600~$^\circ$C  curve does clearly fall apart, which is explained by the fact that at this 
point the recrystallisation is activated, when additionally to the effect of the reduction 
of the dislocation density, an additional effect of the variation of the grain size takes 
place \citep{martinezdeguerenu_jmm07, gurruchaga_metmattrans04b}.

\begin{table*}[ht]
\centering
\begin{tabular}{c|ccccc}
Annealing conditions & $H_c$~(A/m) & $W_h$~(MJ/m$^3$) & $M_r$~(MA/m) & $M_s$~(MA/m) & $\chi_r$ 
\\
\hline
300~$^\circ$C, 51~s & 529.1 & 2901.6 & 0.905 & 1.418 & 636.8 \\
400~$^\circ$C, 51~s & 434.6 & 2479.9 & 1.017 & 1.434 & 673.7 \\
500~$^\circ$C, 51~s & 267.2 & 1875.1 & 1.228 & 1.368 & 896.7 \\
600~$^\circ$C, 51~s & 110.5 & 1199.6 & 1.142 & 1.291 & 984.4 \\
\end{tabular}
\caption{Hysteresis features obtained after fitting to the experimental curves corresponding to 
51~s isothermal annealing at the four different considered temperatures.}
\label{tab:optvalues}
\end{table*}

\section*{Conclusions}

A regressor-based hybridisation of two parametric models has been proposed in order to 
describe sets of experimental curves with a broad range of features. The present work follows 
a procedure, where the physical hysteresis model is replaced by a generic meta-model, a 
similar idea to the one previously exploited in \citep{skarlatos_physicab18}. 

The use of the regressor presents a number of advantages, when a pragmatic approach is sought
in order to reproduce experimental data with a reasonable accuracy. The computationally 
expensive model identification (i.e. the calculation of the model parameters via fitting to 
the experimental curve), is significantly accelerated since each call to the physical model is 
replaced by a regressor evaluation, which is carried out in nearly real time. This 
acceleration, also discussed in \citep{skarlatos_physicab18}, is particularly interesting 
with models like the Jiles-Atherton model, where the evaluation of a curve signifies the 
solution of a differential equation. Furthermore, the above adopted approach stabilises the
identification procedure. This can be understood by pointing the fact that the minimisation 
algorithm (which can be either a deterministic conjugate-gradient-based algorithm or a 
stochastic approach like evolutionary or genetic algorithms) may enter in domains where the 
physical model yields unphysical solutions. This problem is avoided by the selection taking 
place in the off-line phase of the regressor construction. Finally, the fact that the approach 
is not depending on model-specific variables (the regressor is trained using input-output 
pairs) allows the mixing of more than one physical models, and the extension of the domain of 
validity of each one of them.

Although two particular physical models have been considered in this work, namely the 
Jiles-Atherton and the Mel'gui model, the approach is general and can be applied with an 
arbitrary combination of models.



\begin{thebibliography}{10}
\expandafter\ifx\csname url\endcsname\relax
  \def\url#1{\texttt{#1}}\fi
\expandafter\ifx\csname urlprefix\endcsname\relax\def\urlprefix{URL }\fi
\expandafter\ifx\csname href\endcsname\relax
  \def\href#1#2{#2} \def\path#1{#1}\fi

\bibitem{martinezdeguerenu_actamater04a}
A.~{Mart\'inez-de-Guerenu}, F.~Arizti, M.~D\'iaz-Fuentes, I.~Guti\'errez,
  Recovery during annealing in a cold rolled low carbon steel. {Part I}:
  Kinetics and microstructural characterization, Acta Mater. 52~(12) (2004)
  3657--3664.
\newblock \href {https://doi.org/10.1016/j.actamat.2004.04.019}
  {\path{doi:10.1016/j.actamat.2004.04.019}}.

\bibitem{martinezdeguerenu_jmm07}
A.~{Mart\'inez-de-Guerenu}, K.~Gurruchaga, F.~Arizti, Nondestructive
  characterization of recovery and recrystallization in cold rolled low carbon
  steel by magnetic hysteresis loops, J. Mag. Mag. Mater. 316~(2) (2007)
  e842--e845.
\newblock \href {https://doi.org/10.1016/j.jmmm.2007.03.110}
  {\path{doi:10.1016/j.jmmm.2007.03.110}}.

\bibitem{jiles_jmmm86}
D.~C. Jiles, D.~L. Atherton, Theory of ferromagnetic hysteresis, J. Mag. Mag.
  Mater. 61 (1986) 48--60.
\newblock \href {https://doi.org/{10.1016/0304-8853(86)90066-1}}
  {\path{doi:{10.1016/0304-8853(86)90066-1}}}.

\bibitem{melgui_defekt87}
M.~A. {Mel'gui}, Formulas for describing nonlinear and hysteretic properties of
  ferromagnets, Defektoskopiya 11 (1987) 3--10, (Translated from rusian).

\bibitem{salucci_transgeoscirs16}
M.~Salucci, N.~Anselmi, G.~Oliveri, P.~Calmon, R.~Miorelli, C.~Reboud,
  A.~Massa, Real-time {NDT-NDE} through an innovative adaptive partial least
  squares {SVR} inversion approach, {IEEE} Trans. Geosci. Remote Sens. 54~(11)
  (2016) 6818--6832.
\newblock \href {https://doi.org/10.1109/TGRS.2016.2591439}
  {\path{doi:10.1109/TGRS.2016.2591439}}.

\bibitem{ahmed_ndte19}
S.~Ahmed, C.~Reboud, P.-E. Lhuillier, P.~Calmon, R.~Miorelli, An adaptive
  sampling strategy for quasi real time crack characterization on eddy current
  testing signals, NDT \& E Int. 103 (2019) 154--165.
\newblock \href {https://doi.org/10.1016/j.ndteint.2019.02.001}
  {\path{doi:10.1016/j.ndteint.2019.02.001}}.

\bibitem{forrestereng}
A.~Forrester, A.~Sobester, A.~Keane, Engineering design via surrogate
  modelling: a practical guide, John Wiley \& Sons, 2008.

\bibitem{villemonteix_jglobopt08}
J.~Villemonteix, E.~Vazquez, E.~Walter, An informational approach to the global
  optimization of expensive-to-evaluate functions, Journal of Global
  Optimization 44~(4) (2008) 509--534.
\newblock \href {https://doi.org/10.1007/s10898-008-9354-2}
  {\path{doi:10.1007/s10898-008-9354-2}}.

\bibitem{soto_transim09}
M.~Soto, A.~{Mart\'inez-de-Guerenu}, K.~Gurruchaga, F.~Arizti, A completely
  configurable digital system for simultaneous measurements of hysteresis loops
  and barkhausen noise, {IEEE} Trans. Instrum. Meas. 58~(5) (2009) 1746--1755.
\newblock \href {https://doi.org/10.1109/TIM.2009.2014510}
  {\path{doi:10.1109/TIM.2009.2014510}}.

\bibitem{gurruchaga_metmattrans04b}
K.~Gurruchaga, A.~{Mart\'inez-de-Guerenu}, I.~Guti\'errez, Sensitiveness of
  magnetic inductive parameters for the characterization of recovery and
  recrystallization in cold-rolled low-carbon steel, Metall. Mater. Trans. A
  41A (2010) 985-- 993.
\newblock \href {https://doi.org/10.1007/s11661-009-0156-z}
  {\path{doi:10.1007/s11661-009-0156-z}}.

\bibitem{skarlatos_physicab18}
A.~Skarlatos, A.~Mart\'inez-de Guerenu, R.~Miorelli, A.~Lasaosa, C.~Reboud, Use
  of meta-modelling for identification and interpolation of parametric
  hysteresis models applied to the characterization of carbon steels, Physica B
  549 (2018) 122--126.
\newblock \href {https://doi.org/10.1016/j.physb.2017.11.053}
  {\path{doi:10.1016/j.physb.2017.11.053}}.

\end{thebibliography}

\end{document}